\documentclass{article}
\usepackage{dcase2021_techrep,amsmath,graphicx,url,times,booktabs, tabularx}

\title{Optimizing temporal resolution \\ of convolutional recurrent neural networks \\ for sound event detection }

\name{Wim Boes, Hugo Van hamme}
\address{
ESAT, KU Leuven \\
wim.boes@esat.kuleuven.be, hugo.vanhamme@esat.kuleuven.be}

\begin{document}

\ninept
\maketitle

\begin{sloppy}

\begin{abstract}
In this technical report, the systems we submitted for subtask 4 of the DCASE 2021 challenge, regarding sound event detection, are described in detail. These models are closely related to the baseline provided for this problem, as they are essentially convolutional recurrent neural networks trained in a mean teacher setting to deal with the heterogeneous annotation of the supplied data. However, the time resolution of the predictions was adapted to deal with the fact that these systems are evaluated using two intersection-based metrics involving different needs in terms of temporal localization. This was done by optimizing the pooling operations.

For the first of the defined evaluation scenarios, imposing relatively strict requirements on the temporal localization accuracy, our best model achieved a PSDS score of 0.3609 on the validation data. This is only marginally better than the performance obtained by the baseline system (0.342): The amount of pooling in the baseline network already turned out to be optimal, and thus, no substantial changes were made, explaining this result.

For the second evaluation scenario, imposing relatively lax restrictions on the localization accuracy, our best-performing system achieved a PSDS score of 0.7312 on the validation data. This is significantly better than the performance obtained by the baseline model (0.527), which can effectively be attributed to the changes that were applied to the pooling operations of the network.
\end{abstract}


\section{Introduction}
\label{sec:intro}

The fourth subtask of DCASE 2021~\cite{Turpault} is dedicated to sound event detection. The goal of this problem is to localize and classify sound events in given audio recordings. In previous iterations of this problem~\cite{DCASE2018proceedings, DCASE2019proceedings}, models were ranked using a single event-based score. On the contrary, this year, two intersection-based metrics, representing distinct scenarios, are employed. As explained in Section 3, these measures involve varying requirements with respect to temporal localization accuracy. 

The baseline model supplied for the aforementioned problem~\cite{Turpault} is based on the winning system~\cite{dcase2019winner} of subtask 4 of the DCASE 2019 challenge~\cite{DCASE2019proceedings}. It was not designed to deal with sound event detection situations entailing different needs in terms of event localization accuracy. In particular, it is important to point out that the temporal resolution of the output of this system is fixed, which may not be optimal for every possible scenario. 

In this work, we therefore investigate how adapting the pooling operations of the baseline model, which are directly connected to the temporal resolution of its output, impacts the performance of this system under different evaluation setups.

In Section~\ref{sect:models}, the submitted systems are described in detail. In Section~\ref{sect:setup}, we expand upon the experimental setup. Then, in Section~\ref{sect:results}, we report the results obtained by the considered models, and finally, we draw a conclusion in Section~\ref{sect:conclusion}.

\section{Models}
\label{sect:models}

In this section, the submitted models, which are slight adaptations of the baseline for the considered challenge~\cite{Turpault}, are elaborated upon.

\subsection{Architecture}

The input to the system is a spectral map of an audio recording. The number of frequency bins is predefined and set to 128.

The first component of the architecture is a convolutional neural network (CNN), consisting of seven blocks. Each  of those comprises the following five layers: a convolutional layer, a batch normalization layer~\cite{batchnorm}, a ReLU activation layer, a dropout layer~\cite{dropout} (with a drop rate of 33\%) and an average pooling layer. 

Each convolutional layer uses a square kernel of size 3 and employs a stride of 1. The amount of output channels of these operations varies per block: For the first three, 16, 32 and 64 filters are utilized respectively. For the last four, this number is equal to $128$.

The hyperparameters of the pooling operations were optimized per evaluation scenario (explained in Section~\ref{sect:setup}). The resulting kernel sizes and strides (which are identical per block) are given in Table~\ref{tab:poolhyper}. The first and second numbers of each tuple in this list relate to the time and frequency axes respectively.

At the end of the last block, the frequency-related dimension of the input spectral map has been reduced to one and the corresponding axis can consequently be discarded. 

\begin{table}[!ht]
\caption{Kernel sizes and strides of average pooling layers in CNN}
\label{tab:poolhyper}
\centering
\begin{tabular}{@{}lcc@{}}
\toprule
\textbf{Block} & \textbf{Evaluation scenario 1} & \textbf{Evaluation scenario 2} \\
\midrule
0-1 & (2, 2) & (2, 2) \\
2-3-4 & (1, 2) & (2, 2) \\
4-5-6 & (1, 2) & (1, 2) \\
\bottomrule
\end{tabular}
\end{table}

Afterwards, a two-layered bidirectional gated recurrent unit with a hidden size of 128 is used to model temporal dependencies. The output of this layer is followed by a linear map and application of the sigmoid function to obtain multi-label frame-level probabilities. These scores are aggregated to get clip-level audio event probabilities by performing linear (softmax) pooling~\cite{poolcomparison}.

\subsection{Postprocessing}

During evaluation (see Section~\ref{sect:setup}), the frame-level outputs of the network are turned into binary values by using a number of thresholds. After applying each threshold, those decisions are also passed through a median filter with a 7 frame window. For the models built for evaluation scenarios 1 and 2, this corresponds to durations of about 0.5 and 4 seconds respectively. Finally, the binary frame-level output is converted into event predictions by performing a merging operation with a maximum gap tolerance of 200 ms.

\subsection{Mean teacher training principle}

As explained in Section~\ref{sect:setup}, the training data supplied for the considered challenge is heterogeneously annotated: Some of the  provided recordings include weak (clip-level) labels, some come with strong (event-level) labels and the rest is unlabeled. To deal with this, the mean teacher principle~\cite{meanteacher} is employed.

The parameters of the teacher are computed as the exponential moving average of the student model weights with a multiplicative decay factor of 0.999 per training iteration.

The loss used to train the student consists of four components: The first two are clip-level and frame-level binary cross entropy functions, which are calculated for respectively the weakly and strongly annotated data samples only. The two others are mean-squared error consistency costs between the clip-level and frame-level outputs of the student and teacher models. The classification and consistency terms are summed with weights one and two respectively to obtain the final loss. 

\section{Experimental setup}
\label{sect:setup}

\subsection{Data}
\label{sect:data}

The multi-label data set for problem 4 of DCASE 2021~\cite{Turpault} consists of audio files with a maximum length of 10 seconds. There are 10 possible sound event categories, which are not mutually exclusive. The training partition consists of three subsets:
\begin{itemize}
    \item 1578 real recordings with weak, clip-level labels
    \item 14412 real recordings with no labels
    \item 10000 synthetic recordings with strong, event-level labels
\end{itemize}

The validation set holds 1168 strongly labeled real recordings.

\subsection{Preprocessing}

To get spectral maps of the audio recordings, we first resampled all clips to 22.050 kHz and performed peak amplitude normalization. Next, log mel spectrograms with 128 frequency bins were extracted using a window size of 2048 samples and a hop length of 363 samples. Lastly, per-frequency bin standardization was carried out.

\subsection{Data augmentation}

In order to avoid the risk of overfitting, we employed several types of data augmentation during the training of the considered models.

Firstly, we used time and frequency masking~\cite{specaugment}. The maximum lengths of the time and frequency masks applied to the spectral input maps were equal to 25/100 and 32/16 respectively for the models built for evaluation scenario 1/2.

Secondly, we employed mixup~\cite{mixup} with a probability of 50\% of applying it. The mixing ratios were randomly sampled from a beta distribution with shape parameters equal to 0.2.

\subsection{Training and evaluation}

All of the models were trained and evaluated using PyTorch~\cite{pytorch}.

\subsubsection{Training hyperparameters}

All systems were trained for 200 epochs. Per epoch, 250 batches of 48 samples were given to the networks. Each batch contained 12 weakly labeled, 12 strongly annotated and 24 unlabeled examples.

Adam~\cite{adam} was employed to train the weights of the student models. Learning rates were ramped up exponentially from 0 to 0.001 for the first 12500 optimization iterations. Subsequently, they decayed multiplicatively at a rate of 0.99995 per training step.

\subsubsection{Evaluation scenarios}

For this challenge, there are two evaluation scenarios. The first of those imposes relatively strict requirements on the temporal localization accuracy, the second is much more lax in this regard. 

In both scenarios, the PSDS score~\cite{psds} is used as measure, but the exact hyperparameters (listed in Table~\ref{tab:psdshyper})  vary per situation.

\begin{table}[!ht]
\caption{PSDS hyperparameters per evaluation scenario}
\label{tab:psdshyper}
\centering
\begin{tabular}{@{}lcc@{}}
\toprule
\textbf{Hyperparameter} & \textbf{Scenario 1} & \textbf{Scenario 2} \\
\midrule 
Detection tolerance criterion & 0.7 & 0.1 \\
Ground truth intersection criterion & 0.7 & 0.1 \\
Cross-trigger tolerance criterion & N/A & 0.3 \\
Cost of class instability & 1 & 1 \\
Cost of cross-triggers & 0 & 0.5 \\
Maximum false positive rate & 100 & 100 \\ 
\bottomrule
\end{tabular}
\end{table}

All PSDS scores were calculated using the probabilities of the students after the last training epoch.

\section{Results}
\label{sect:results}

For the first evaluation setting, the baseline obtained a PSDS value of 0.342 on the validation data.
Our best model achieved a score of 0.3609, which is only marginally better. This minor improvement can only be explained by the amplified data augmentation, not by a difference in temporal resolution between the two systems: We found the amount of pooling present in the baseline model to already be optimal for this particular scenario, and therefore, no changes were made in this regard. 

For the second evaluation setting, the baseline obtained a PSDS value of 0.527 on the validation data.
Our best model achieved a score of 0.7312, which is a major improvement. In this case, the changed amount of pooling - and therefore the difference in temporal resolution of the systems - plays the most significant role. 
\section{Conclusion}
\label{sect:conclusion}

In this technical report, we described the systems we submitted for task 4 of the DCASE 2021 challenge and reported their results. 

The best PSDS scores for the two predefined evaluation scenarios achieved by our models were 0.3609 and 0.7312 respectively.

\section{Acknowledgment}
\label{sec:ack}

This work is supported by a PhD Fellowship of Research Foundation Flanders (FWO-Vlaanderen).

\bibliographystyle{IEEEtran}
\bibliography{refs}

\begin{thebibliography}{10}
\providecommand{\url}[1]{#1}
\def\UrlFont{\rmfamily}
\providecommand{\newblock}{\relax}
\providecommand{\bibinfo}[2]{#2}
\providecommand\BIBentrySTDinterwordspacing{\spaceskip=0pt\relax}
\providecommand\BIBentryALTinterwordstretchfactor{4}
\providecommand\BIBentryALTinterwordspacing{\spaceskip=\fontdimen2\font plus
\BIBentryALTinterwordstretchfactor\fontdimen3\font minus
  \fontdimen4\font\relax}
\providecommand\BIBforeignlanguage[2]{{%
\expandafter\ifx\csname l@#1\endcsname\relax
\typeout{** WARNING: IEEEtran.bst: No hyphenation pattern has been}%
\typeout{** loaded for the language `#1'. Using the pattern for}%
\typeout{** the default language instead.}%
\else
\language=\csname l@#1\endcsname
\fi
#2}}

\bibitem{Turpault}
N.~Turpault, R.~Serizel, A.~Parag~Shah, and J.~Salamon, ``{Sound event
  detection in domestic environments with weakly labeled data and soundscape
  synthesis},'' in \emph{{Workshop on Detection and Classification of Acoustic
  Scenes and Events}}, 2019.

\bibitem{DCASE2018proceedings}
M.~D. Plumbley, C.~Kroos, J.~P. Bello, G.~Richard, D.~P. Ellis, and A.~Mesaros,
  \emph{{Proceedings of the Detection and Classification of Acoustic Scenes and
  Events 2018 Workshop (DCASE2018)}}, 2018.

\bibitem{DCASE2019proceedings}
M.~Mandel, J.~Salamon, and D.~P. Ellis, \emph{{Proceedings of the Detection and
  Classification of Acoustic Scenes and Events 2019 Workshop (DCASE2019)}},
  2019.

\bibitem{dcase2019winner}
L.~Delphin-Poulat and C.~Plapous, ``{Mean teacher with data augmentation for
  dcase 2019 task 4},'' Orange Labs Lannion, Tech. Rep., 2019.

\bibitem{batchnorm}
S.~Ioffe and C.~Szegedy, ``{Batch Normalization: Accelerating Deep Network
  Training by Reducing Internal Covariate Shift},'' in \emph{Proceedings of
  ICML}, 2015, pp. 448--456.

\bibitem{dropout}
N.~Srivastava, G.~Hinton, A.~Krizhevsky, I.~Sutskever, and R.~Salakhutdinov,
  ``{Dropout: A Simple Way to Prevent Neural Networks from Overfitting},''
  \emph{The Journal of Machine Learning Research}, vol.~15, no.~1, pp.
  1929--1958, 2014.

\bibitem{poolcomparison}
Y.~Wang, J.~Li, and F.~Metze, ``A comparison of five multiple instance learning
  pooling functions for sound event detection with weak labeling,'' in
  \emph{2019 IEEE International Conference on Acoustics, Speech and Signal
  Processing (ICASSP)}, 2019, pp. 31--35.

\bibitem{meanteacher}
A.~Tarvainen and H.~Valpola, ``{Mean teachers are better role models:
  Weight-averaged consistency targets improve semi-supervised deep learning
  results},'' in \emph{Advances in Neural Information Processing Systems},
  2017, pp. 1195--1204.

\bibitem{specaugment}
D.~S. Park, W.~Chan, Y.~Zhang, C.-C. Chiu, B.~Zoph, E.~D. Cubuk, and Q.~V. Le,
  ``{SpecAugment: A simple data augmentation method for automatic speech
  recognition},'' \emph{arXiv preprint arXiv:1904.08779}, 2019.

\bibitem{mixup}
H.~Zhang, M.~Cisse, Y.~N. Dauphin, and D.~Lopez-Paz, ``{mixup: Beyond empirical
  risk minimization},'' \emph{arXiv preprint arXiv:1710.09412}, 2017.

\bibitem{pytorch}
A.~Paszke, S.~Gross, F.~Massa, A.~Lerer, J.~Bradbury, G.~Chanan, T.~Killeen,
  Z.~Lin, N.~Gimelshein, L.~Antiga, A.~Desmaison, A.~Kopf, E.~Yang, Z.~DeVito,
  M.~Raison, A.~Tejani, S.~Chilamkurthy, B.~Steiner, L.~Fang, J.~Bai, and
  S.~Chintala, ``{PyTorch: An Imperative Style, High-Performance Deep Learning
  Library},'' in \emph{Advances in Neural Information Processing Systems},
  2019, pp. 8024--8035.

\bibitem{adam}
D.~P. Kingma and J.~Ba, ``{Adam: A Method for Stochastic Optimization},''
  \emph{arXiv preprint arXiv:1412.6980}, 2014.

\bibitem{psds}
{\c{C}}.~Bilen, G.~Ferroni, F.~Tuveri, J.~Azcarreta, and S.~Krstulovi{\'c},
  ``{A framework for the robust evaluation of sound event detection},'' in
  \emph{2020 IEEE International Conference on Acoustics, Speech and Signal
  Processing (ICASSP)}, 2020, pp. 61--65.

\end{thebibliography}

\end{sloppy}
\end{document}